\documentclass[]{spie}  

\usepackage{amsmath,amsfonts,amssymb}
\usepackage{graphicx}
\usepackage[colorlinks=true, allcolors=blue]{hyperref}
\usepackage{astro_bib_macro} 

\title{High-Contrast Imager for Complex Aperture Telescopes (HiCAT): 5. First Results With Segmented-Aperture Coronagraph and Wavefront Control}

\author{R\'emi Soummer\supit{a}, 
Gregory R. Brady\supit{a}, 
Keira Brooks\supit{a}, 
Thomas Comeau\supit{a}, 
\'Elodie Choquet\supit{e}, 
Tom Dillon \supit{n},
Sylvain Egron\supit{g}, 
Rob Gontrum\supit{a}, 
John Hagopian\supit{h},
Iva Laginja\supit{a}, 
Lucie Leboulleux\supit{a}\supit{b}\supit{c}, 
Marshall D. Perrin\supit{a}, 
Peter Petrone\supit{a}, \supit{m}
Laurent Pueyo\supit{a}, 
Johan Mazoyer\supit{k}, 
Mamadou N'Diaye\supit{d}, 
A.J. Eldorado Riggs \supit{f},
Ron Shiri\supit{j},
Anand Sivaramakrishnan\supit{a}, 
Kathryn St. Laurent\supit{a}, 
Ana-Maria Valenzuela\supit{a},  
Neil T. Zimmerman\supit{j},
\skiplinehalf
\supit{a} Space Telescope Science Institute, 3700 San Martin Drive, Baltimore, MD 21218, USA\\
\supit{b} Laboratoire d\'{}Astrophysique de Marseille UMR 7326, 13388, Marseille, France\\
\supit{c} Office National d\'{}Etudes et de Recherches A\'{e}rospatiales, 92320 Ch\^{a}tillon, France\\
\supit{d} Laboratoire Lagrange, Observatoire de Nice C\^{o}te d\'{}Azur, 06304 Nice, France\\
\supit{e} California Institute of Technology,  MC 249-17, Pasadena, CA 91125, USA\\
\supit{f} Jet Propulsion Laboratory, 4800 Oak Grove Drive, Pasadena, CA 91109, USA\\
\supit{g} Iridescence S.A.R.L., 149 avenue du Maine, 75014 Paris, France\\
\supit{h} Advanced Nanophotonics Inc., 4437 Windsor Farm Rd,
Harwood, MD USA\\
\supit{i} Iridescence S.A.R.L., 149 avenue du Maine, 75014 Paris, France\\
\supit{j} NASA Goddard Space Center, 8800 Greenbelt Rd, Greenbelt, MD 20771, USA\\
\supit{k} John Hopkins University, 3400 North Charles Street, Baltimore, MD 21218, USA\\
\supit{m} Sigma Space Corporation, 4600 Forbes Blvd, Lanham, MD  20706, USA\\
\supit{n} University of Delaware, Newark, DE 19716 USA\\
}

\authorinfo{Further author information, send correspondence to R\'emi Soummer: E-mail: soummer@stsci.edu}

\begin{document} 
\maketitle

\begin{abstract}

Segmented telescopes are a possible approach to enable large-aperture space telescopes for the direct imaging and spectroscopy of habitable worlds. However, the increased complexity of their aperture geometry, due to the central obstruction, support structures and segment gaps, makes high-contrast imaging very challenging. The High-contrast Imager for Complex Aperture Telescopes (HiCAT) testbed was designed to study and develop solutions for such telescope pupils using wavefront control and coronagraphic starlight suppression. The testbed design has the flexibility to enable studies with increasing complexity for telescope aperture geometries starting with off-axis telescopes, then on-axis telescopes with central obstruction and support structures - e.g. the Wide Field Infrared Survey Telescope (WFIRST) - up to on-axis segmented telescopes, including various concepts for a Large UV, Optical, IR telescope (LUVOIR).  In the past year, HiCAT has made significant hardware and software updates in order to accelerate the development of the project. In addition to completely overhauling the software that runs the testbed, we have completed several hardware upgrades, including the second and third deformable mirror, and the first custom Apodized Pupil Lyot Coronagraph (APLC) optimized for the HiCAT aperture, which is similar to one of the possible geometries considered for LUVOIR.   The testbed also includes several external metrology features for rapid replacement of parts, and in particular the ability to test multiple apodizers readily, an active tip-tilt control system to compensate for local vibration and air turbulence in the enclosure.  On the software and operations side, the software infrastructure enables 24/7 automated experiments that include routine calibration tasks and high-contrast experiments. In this communication we present an overview and status update of the project, both on the hardware and software side, and describe the results obtained with APLC wavefront control. 

\end{abstract}


\section{INTRODUCTION}
\label{sec:INTRODUCTION}

The discovery and study of worlds that could potentially support life has emerged as perhaps the most compelling science driver for future large space observatories. A ``New Worlds Mission'' that obtains direct spectroscopic measurements of terrestrial planet atmospheres would enable us to search for signatures of habitability and seek out potential molecular markers of distant biologies.  Indeed, three of the four large mission concept studies for the 2020 decadal survey have exoplanets as a central driving science goal: the Large UV/Optical/Infrared Surveyor (LUVOIR), the Habitable Exoplanets Imaging Mission (HabEx), and the Origins Space Telescope (OST). The spectroscopic characterization of terrestrial exoplanets is both scientifically profound and potentially transformative for humanity's understanding of our place in the cosmos, and is also a topic of tremendous public interest within the US and worldwide. NASA and National Academies strategic planning committees have consistently recognized this, as seen in the  2010 Decadal Survey and 2016 Midterm Assessment reports, the 2014 and 2018 NASA Strategic Plans, NASA's 2013 Astrophysics Roadmap and 2014 Science Plan, and more. 

An observatory's capability for imaging rocky exoplanets turns out to be a steep function of primary mirror diameter, driving us toward the largest possible apertures \cite{2014ApJ...795..122S,stark2015ApJ}. Building on JWST's technologies, segmented mirror architectures can break through the hard limits that launch vehicle fairings place on the sizes of monolithic mirrors \cite{2014SPIE.9143E..16F}. LUVOIR has baselined a 15-m telescope, with a 9-m alternate, both segmented, and the latter possibly off-axis. HabEx is studying a 6.5 m segmented telescope as its alternative to its 4 m monolith primary architecture. When calculating exo-Earth "yields", or the number of rocky planets in the habitable zone that a mission can detect in a fixed time, these studies usually assume nominal starlight suppression to 10$^{-10}$, which translates into yields of a few to a few dozen exo-Earth candidates for apertures of 4$-$15 m. Achieving coronagraphy at this demanding level of performance on a large segmented aperture will enable discovery of dozens of terrestrial planets and robust atmospheric characterizations in only a few years of total mission time. 

Starlight suppression systems have made substantial progress in the past few years, especially with the WFIRST CGI technology program reaching Technology Readiness Level (TRL) 5 for the coronagraph.   \cite{ilya2014SPIE,goullioud2014SPIE,2016JATIS...2a1003K,2017SPIE10400E..0EC,2016JATIS...2a1004C,2017SPIE10400E..0FS,2016JATIS...2a1019S,2017SPIE10400E..0DS}. The further development and flight of CGI would raise these key technologies to TRL 9, and could begin the possible characterization of Jovian planets in reflected light --- thereby playing an essential role in preparing for a ``search for life'' flagship mission. WFIRST's 2.4 m primary remains too small for a ``search for life'' mission and will not address the technical challenges that are specific to segmented apertures. Even if paired with a ``starshade rendezvous'' \cite{Seager2018AAS...23112109S}, WFIRST could potentially image, at most, a few exo-Earths around the nearest handful of stars. 

Coronagraphy for segmented apertures is necessary to ensure robust options for future missions with the largest possible telescope diameters and the greatest predicted yields of characterized exoplanets. 
Yet, segmented architectures also bring new challenges at both the component and system levels. Coronagraph masks must be carefully optimized for segmented pupils. Segment control must be integrated to mitigate drifts along with the high contrast wavefront control and line-of-sight pointing control systems, leading to a web of interconnected control loops communicating on different timescales. A complex trade space exists between the performance of individual subsystems, for instance trading between passive stability and active control, or between coronagraph throughput, inner working angle, and sensitivity to misalignments\cite{leboulleux2018}. 

Addressing these challenges is the central theme of the HiCAT testbed (High contrast imager for Complex Aperture Telescopes \cite{ndiaye2013SPIE,ndiaye2014SPIE_hicat2,2015SPIE.9605E..0IN,2016SPIE.9904E..3CL,2017SPIE10562E..2ZL}). Starting in 2013, HiCAT has led the way in developing coronagraph and wavefront control solutions for segmented apertures. Recently, other teams have begun turning efforts towards segmented mirrors (e.g.\ at Caltech, Universit\'e de Nice, NASA Ames, GSFC, and JPL; \cite{2017arXiv171202042R,2014ApJ...780..171G, 2016JATIS...2a1018P,2016SPIE.9912E..6LN}). 
HiCAT was designed to address system-level experiments in ambient conditions, in order to pave the way for higher TRL demonstrations in vacuum. This development plan, directly parallel to the WFIRST CGI staged ambient and vacuum experiments, will mature these key technologies needed for future NASA observatories of unprecedented size and power to seek out signs of life on distant rocky worlds.

\section{Testbed Hardware Overview}

\begin{figure}[th!]
\includegraphics[width=1.0\textwidth]{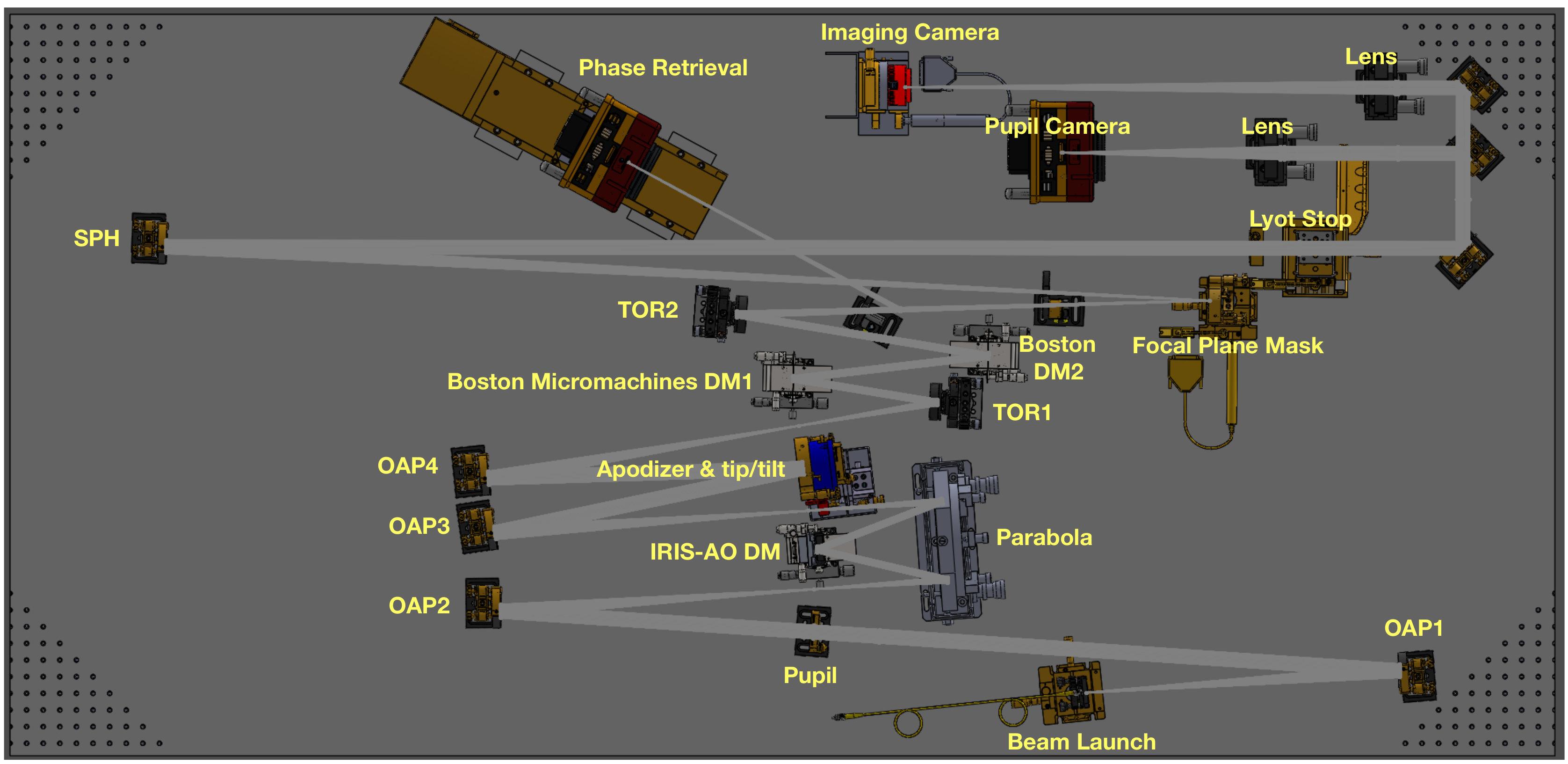}
\caption{HiCAT testbed layout. The testbed includes a telescope simulator (IRIS-AO 37 segment deformable mirror combined with a pupil mask to produce the central obstruction and support structures, an Apodized Pupil Lyot Coronagraph (APLC) combining and apodizer, a hard-edge focal plane mask and a Lyot Stop. Imaging camera include a coronagraphic focal plane, a pupil viewing camera, and a phase retrieval arm to measure the wavefront at the focal plane mask using a high-quality removable fold mirror to minimize non-common path errors. 
\label{fig:hicat_cad} }
\end{figure}

\begin{figure}[!ht]
\includegraphics[width=1\textwidth]{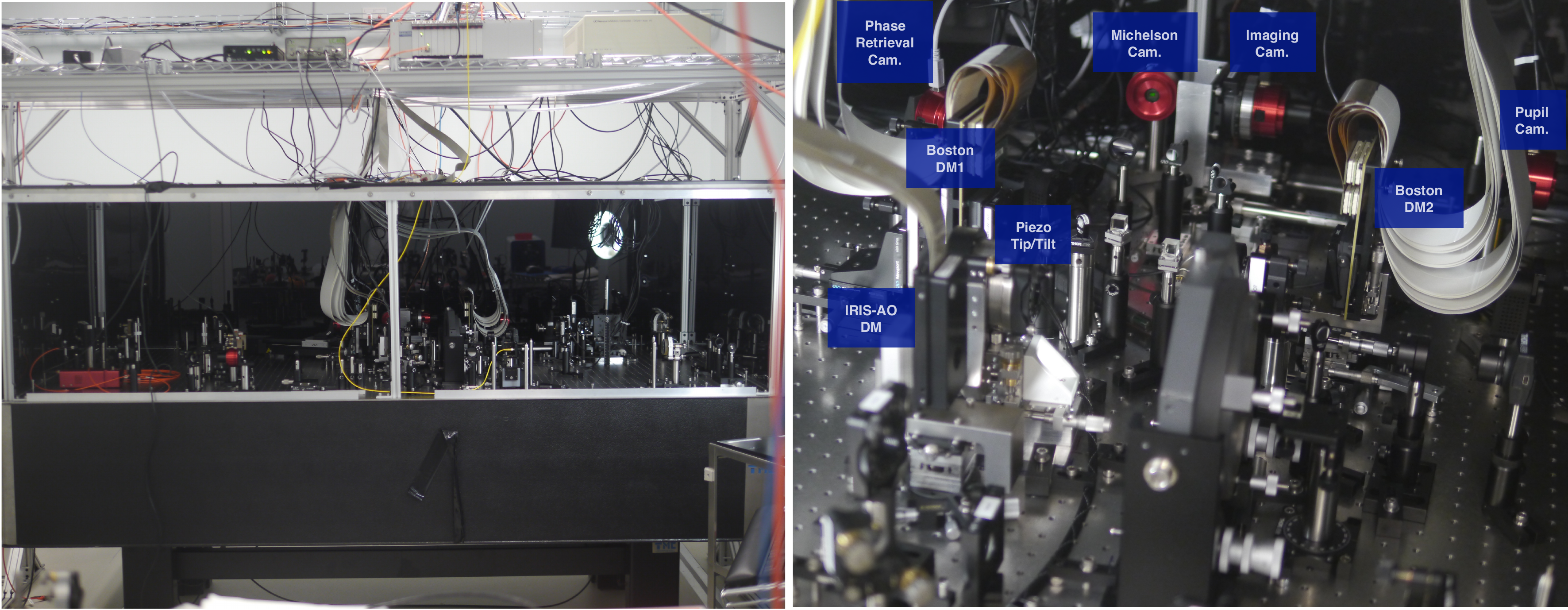}
\caption{Left: view of the entire HiCAT testbed in its (opened) enclosure. Right: detail of the central part of the testbed. Note the metrology channels (Michelson interferometer beamsplitter cubes visible to the right of the piezo tip/tilt label).  These Michelson interferometers are used to enable fast replacement of key components (deformable mirrors, apodizers) under external metrology.}
\label{fig:hicat_pics}
\end{figure}

\begin{figure}[th!]
\includegraphics[width=0.75\textwidth]{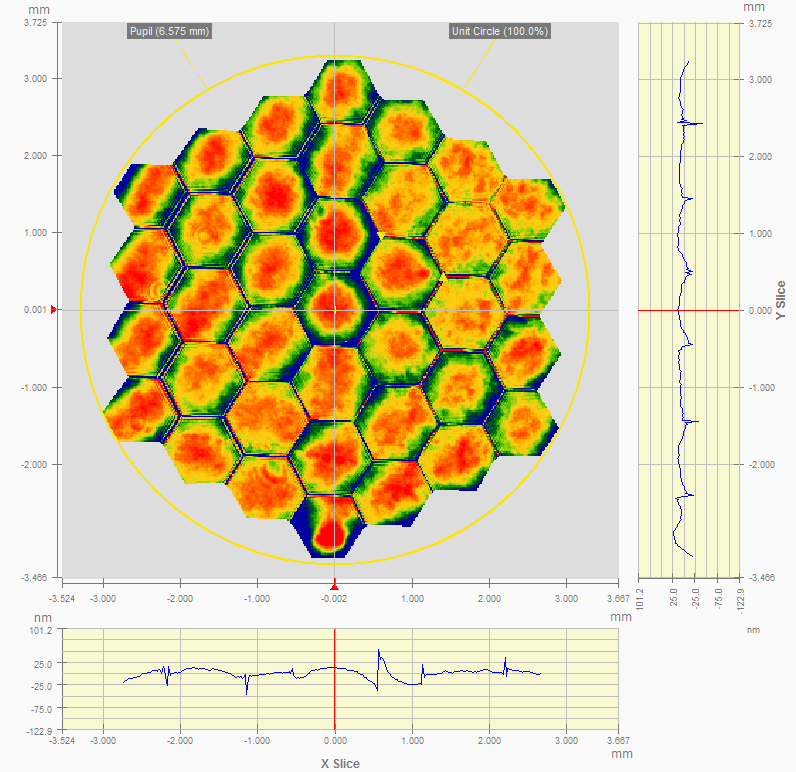}
\caption{Open-loop calibration map of the IRIS-AO ``PTT111L'' 37 segments DM. The segment size is 1.4 mm, and each segment can be controlled in piston, tip and tilt. The vendor-provided calibration corresponds to an horizontal position for the mirror. When used vertically in our testbed, we ground that gravity sag on each segments leads to a ``Venetian blind'' wavefront shape, which we corrected using a Fizeau interferometer\cite{laginja2018}. The final calibrated open-loop flat is shown here, with a surface error of 9nm rms. 
\label{fig:irisao} }
\end{figure}

\begin{figure}[th!]
\includegraphics[width=1.0\textwidth]{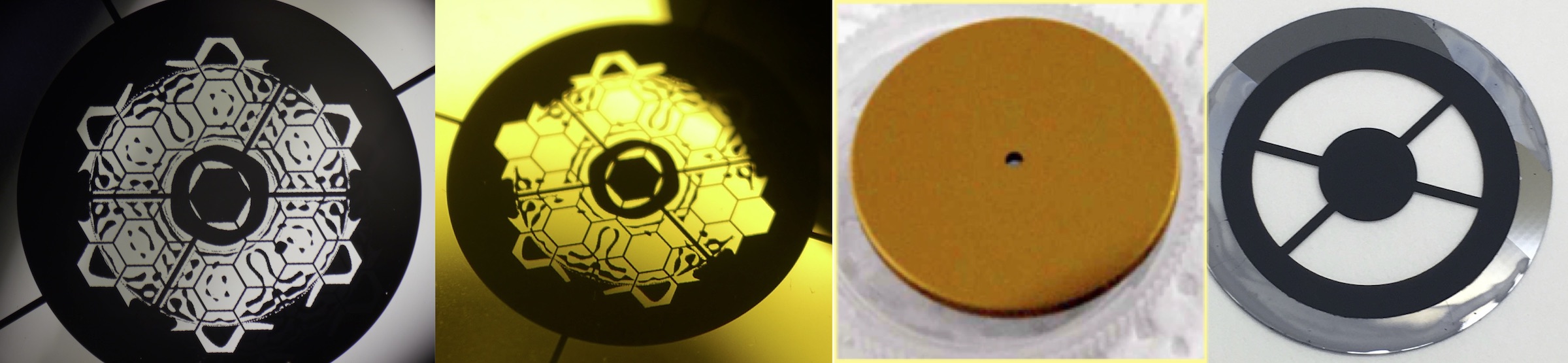}
\caption{\small \textit{Left 2 panels:} First carbon nanotubes apodizers (manufactured by Advanced Nanophotonics Inc.) for an on-axis segmented aperture, on silver and gold. Note the slightly different patterns with some fully reflective segments in the gold case. Recent wavefront control algorithms (ACAD-OSM  \cite{2018AJ....155....7M,2018AJ....155....8M}) will be tested on this ``modified'' apodizer pattern. \textit{Panel 3:} Reflective focal plane mask for the APLC design, re-used from former ``Lyot Project'' courtesy R. Oppenheimer\cite{2004SPIE.5490..433O}. The hole allows natural implementation of a tip/tilt sensor or low-order wavefront sensor\cite{2006SPIE.6272E..2LW}; mask \textit{Panel 4:} Corresponding Lyot Stop with x-shaped support structures, also coated with carbon nanotubes on a through-etched silicon wafer.}
\label{fig:apodizers}
\end{figure}

\begin{figure}[th!]
\includegraphics[width=0.6\textwidth]{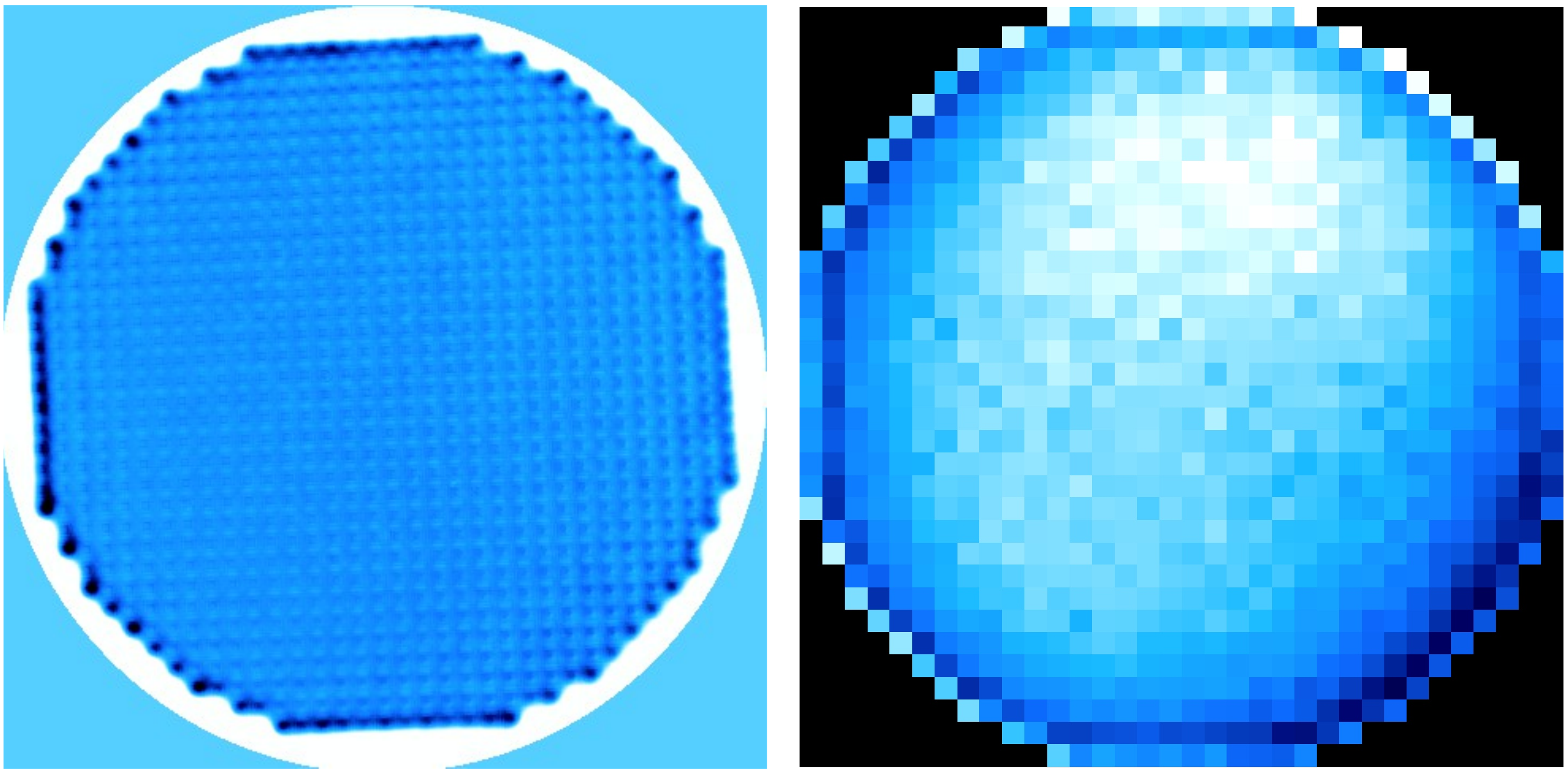}
\caption{Our Boston Micromachines ``kilo-DM'' was custom calibrated in a closed loop using our Fizeau interferometer. This process allows to generate not only flat calibration files (left) but also a library of well-calibrated shapes (zernikes polynomials of various amplitudes, sine waves, wavefront sensing probes) that can be used either for wavefront sensing or sensitivity analyses. Left: closed-loop calibrated flat map reaching the accuracy limits of our Fizeau interferometer, surface error $\sim1-2$ nm rms. The fine structure corresponds to actuator print-through. Right: corresponding voltage map applied to the DM to produce the flat map. 
\label{fig:BMM} }
\end{figure}

The HiCAT testbed project started in 2013. The goal of the testbed is to demonstrate coronagraphy and wavefront control with an on-axis segmented aperture. The testbed layout and hardware are illustrated in Figure \ref{fig:hicat_cad} and \ref{fig:hicat_pics}. The testbed includes the following elements: 

\begin{itemize}
    \item A segmented telescope simulator with central obstruction and support structures using a 37-segment, aperture DM (IRIS-AO\footnote{http://www.irisao.com}). The central obstruction and support structures can be added in the first pupil plane, using a laser-cut mask. This truly-segmented telescope simulator includes real co-phasing wavefront errors and control ability, with an open-loop calibrated surface error of 9 nm rms (see Figure \ref{fig:irisao}). The repeatability of the open-loop calibration at this level of wavefront quality makes it directly suitable for the high-contrast goals of HiCAT.   
   
    \item A custom Apodized Pupil Lyot Coronagraph \cite{ndiaye2015ApJ_aplc4, ndiaye2016} including a shaped pupil apodizer manufactured by Advanced Nanophotonics Inc.  \cite{2010SPIE.7761E..0FH} using carbon nanotubes coatings, a circular hard-edge focal plane mask re-using a mirror with an ultra-precise circular hole originally from the Lyot Project, which was the first extreme adaptive optics coronagraph\cite{2004SPIE.5490..433O} (courtesy Rebecca Oppenheimer, AMNH); and a silicon-etched Lyot stop with carbon-nanotube coating as well (see Figure \ref{fig:apodizers} for detailed pictures of that hardware).
    
    \item Two Boston Micromachines \footnote{http://www.bostonmicromachines.com} ``kilo" DMs (932 actuators) are used for wavefront control (one in pupil plane and one out of pupil plane) and were custom-calibrated in closed-loop using our Fizeau interferometer. The residual flat calibration is within the accuracy limits of our 4D\footnote{https://www.4dtechnology.com} Fizeau interferometer and the residual surface error is of the order of 1-2 nm rms (Figure \ref{fig:BMM}).
    
    \item An optional flat mirror can be introduced in the beam to redirect the beam to a phase retrieval camera on a motorized translation stage to analyze the wavefront at the focal plane mask using a parametric phase retrieval algorithm\cite{brady2018}.
    
    \item Several low-cost CMOS cameras are installed on HiCAT (pupil and focal plane imaging, phase retrieval, and dedicated cameras for our each of our Michelson Interferometers used for metrology.  These CMOS camera have all been  upgraded in the past year and feature very small pixels, very low noise ($\sim2$ electrons), and fast readout modes\footnote{https://astronomy-imaging-camera.com}, which facilitate the HiCAT operations. 
    
    \item Testbed motorization includes XYZ translation for the focal plane mask and Lyot stop, and z-translation stages for the imaging camera and phase retrieval stage.
    
    \item Several light sources options are available to the testbed, including a laser diode at 638nm, a super-continuum laser, and a Photon ETC\footnote{http://www.photonetc.com} tunable filter that delivers narrow band adjustable 1\% bandpass light. For higher flux density we have built a custom light source assembly using motorized filter wheels and interference filters.
    
\end{itemize}

\section{Current Results and Achievements}
\subsection{Full-system dark zone demonstration}

\begin{figure}[th!]
\includegraphics[width=1.0\textwidth]{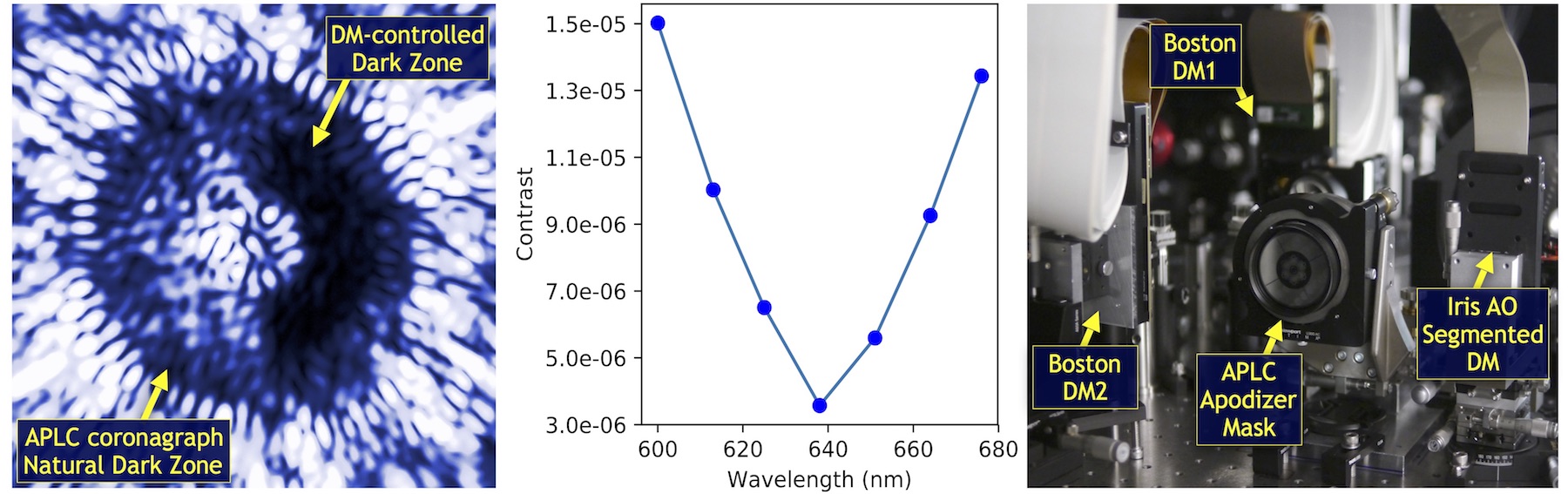}
\caption{\small \textit{Left:} First demonstration of a coronagraph dark zone with a segmented aperture on-axis telescope simulator (monochromatic contrast: $1.7\times10^{-6}$, broadband contrast $6.3\times10^{-6}$ in $6\%$ bandpass).  Note the ``natural'' circular dark zone of the APLC, and the deeper dark zone on the right from wavefront control. \textit{Center:} Broadband demo using a tunable laser (not yet fully optimized in this experiment, with minimum contrast about 2x higher than our best monochromatic result). \textit{Right:} HiCAT testbed with three DMs and carbon nanotubes APLC apodizer. \textit{\textbf{Note:} These results were obtained upon installation of the segmented DM and controlling just one continuous DM (others set to flat); it does not represent the limits of the testbed.} 
\label{fig:hicat_results} }
\end{figure}

The testbed hardware and software main infrastructure was completed in the past year, in particular with the installation and calibration of the three deformable mirrors and the implementation of the first segmented-aperture APLC.  These efforts have culminated in the first-ever high-contrast dark zone with a segmented aperture coronagraph operating on an actual segmented pupil, achieving initial contrasts of $1.7\times10^{-6}$ monochromatic and $6.3\times10^{-6}$ in $6\%$ broadband with our latest APLC design operating in air (see Figure \ref{fig:hicat_results}). This major milestone in  high-contrast coronagraphy for segmented apertures establishes with high confidence that the shaped pupil APLC has reached TRL 3.

\subsection{APLC coronagraph design and manufacturing}

The Apodized Pupil Lyot Coronagraph\cite{ASF02,SAF03,2005ApJ...618L.161S,2009ApJ...695..695S,2011ApJ...729..144S} was extended in the past few years to work with shaped-pupil apodizations, and specifically optimized to the configuration of on-axis segmented telescopes\cite{ndiaye2015ApJ_aplc4, ndiaye2016,zimmerman2016jatis}.  
Our APLC solution relies on similar reflective binary shaped-pupil masks as those used in WFIRST-CGI and can therefore build directly on mature, proven technology for component manufacturing, for example using JPL's silicon grass technology\cite{2017SPIE10400E..0CB}. We have been developing an alternate manufacturing option for shaped-pupil apodizers, in collaboration with Advanced Nanophotonics Inc. \cite{2010SPIE.7761E..0FH}.  This uses carbon nanotubes grown on a customized catalyst for the black areas, and protected silver or gold for the reflective areas (see Figure \ref{fig:apodizers}).
The hemispherical reflectance of carbon nanotubes on these apodizers is $0.5\%$. Our next batch will target reflectance below $0.1\%$, comparable or better than JPL's silicon grass used for the WFIRST CGI masks. In theory, carbon nanotubes can reach hemispherical reflectance down to about 0.05\%. They offer potentially better surface figure because they involve growth on a catalyst instead of surface etching. While still a research topic, catalyst customization could be used to enable non binary masks with gray areas. There are no limitations for the reflective surface (e.g. silver and gold, as shown in figure \ref{fig:apodizers}). 
We plan to implement both a custom JPL-manufactured black silicon mask (in the coming year) as well as continue investigating carbon nanotubes coatings, and will compare their performance and respective advantages.  

\begin{figure}[th!]
\includegraphics[width=1.0\textwidth]{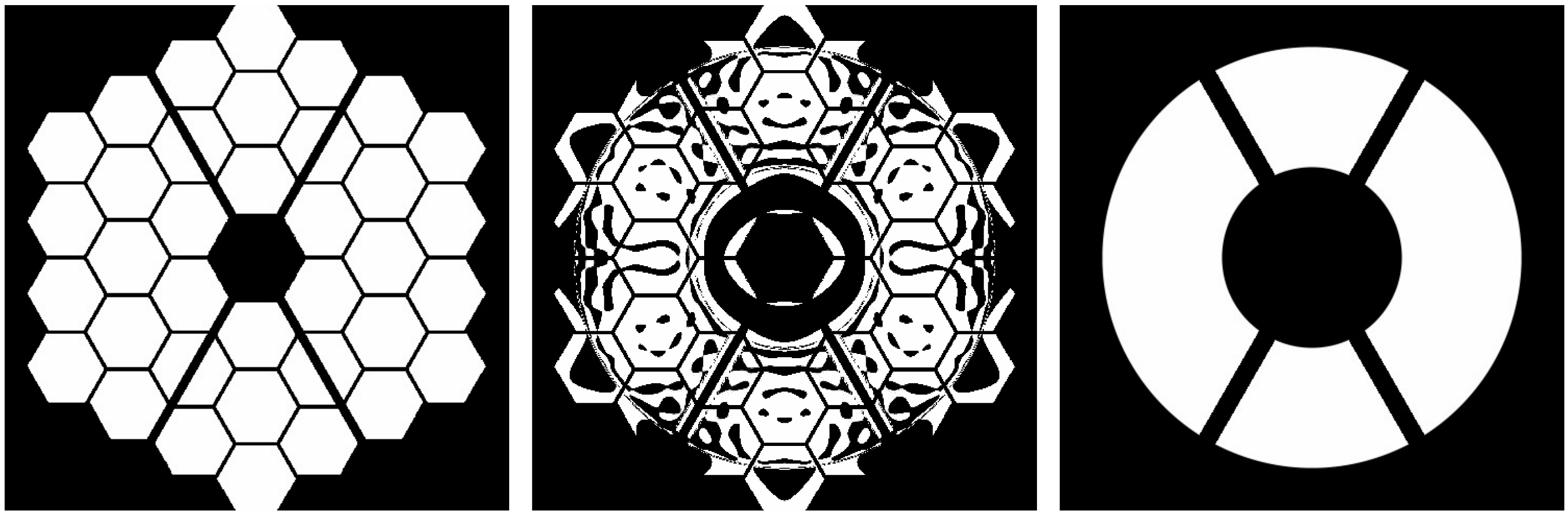}
\caption{HiCAT APLC geometry (left: pupil geometry with central obstruction, support structures and segment gaps).  Segment gaps are oversized compared to the natural gaps of the IRIS-AO deformable mirror, and the perimeter has been slightly undersized to accommodate for alignment tolerances. The X-shaped support structure on this 37-segment aperture is one of the plausible designs for a telescope like LUVOIR that was initially established by the NASA Exoplanet Exploration Program (ExEP) Segmented-aperture Coronagraph Design and Analysis (SCDA)\cite{2016SPIE.9904E..1YZ,2017SPIE10398E..0HC,st.laurent2018}.
\label{fig:aplc_geometry} }
\end{figure}

\begin{figure}[th!]
\includegraphics[width=0.75\textwidth]{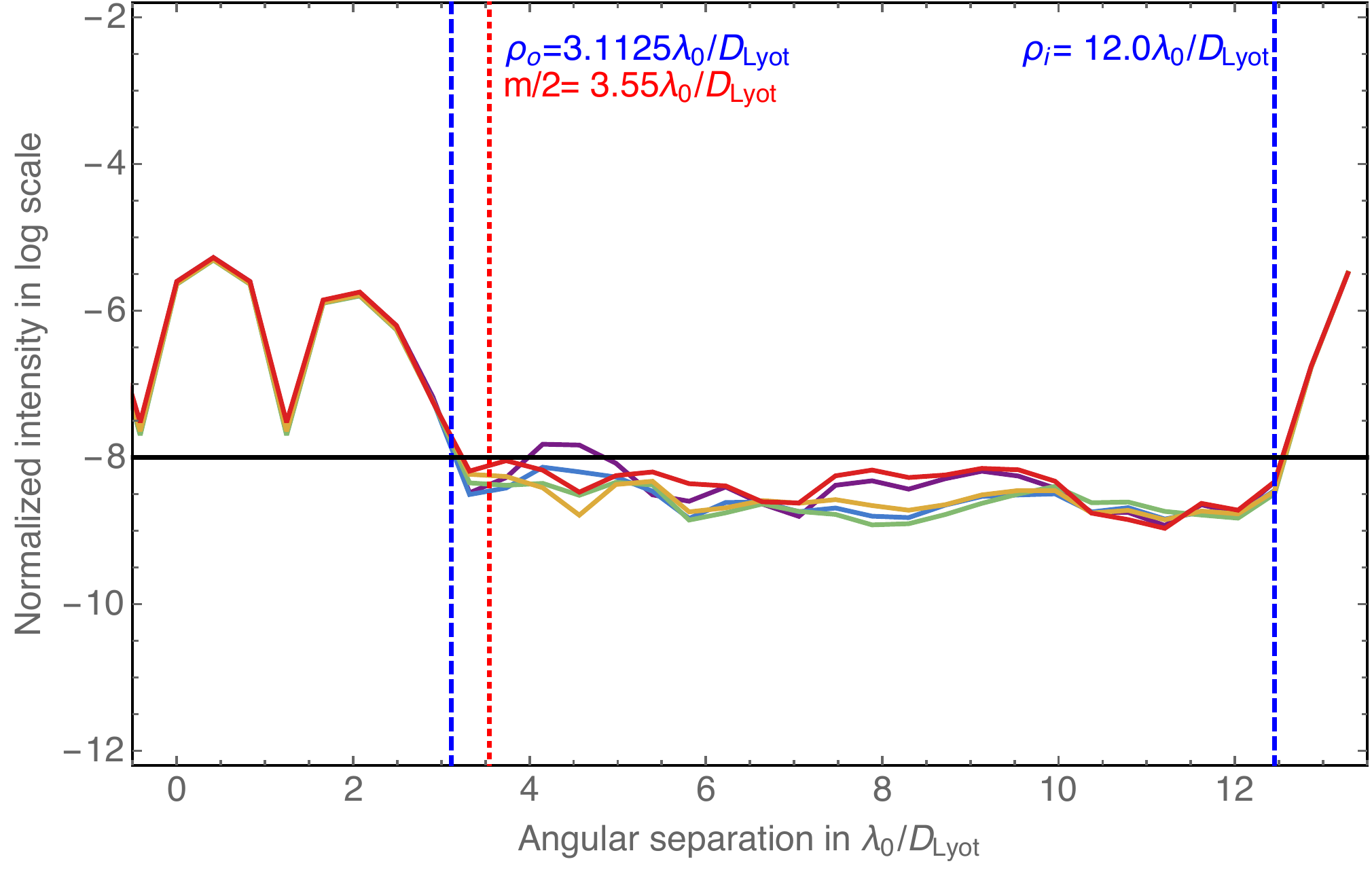}
\caption{APLC radial contrast for a 10\% bandpass (design-limited). The dashed blue line indicate the natural IWA/OWA of the APLC design, the OWA is chosen to match the controllable region of the DM, and the APLC IWA was designed to be smaller than the focal plane mask radius for increased robustness to low-order aberrations\cite{ndiaye2015ApJ_aplc4, ndiaye2016}. Each color corresponds to the contrast profile for a misaligned Lyot stop position (here by 0.1\% of the pupil diameter). More robust solutions in the range 0.6--1.0\% are reachable and will be implemented in future apodizer manufacturing runs. 
\label{fig:aplc_contrast_profile} }
\end{figure}

The APLC design is optimized for an aperture with  geometric features slightly oversized/undersized to account for alignment tolerancing (Figure \ref{fig:aplc_geometry}).  The selected FPM mask diameter is $8.543 \lambda/D_{apod}$ based on existing hardware (Figure \ref{fig:apodizers}). The APLC dark zone is optimized for an IWA of 3.75 to 15 $\lambda/D_{apod}$.  Based on our projected aperture on the DM, our maximum controllable spatial frequency is $14.9\lambda/D_{apod}$. Note that the APLC IWA ($3.75\lambda/D_{apod}$) is smaller than the projected radius of the focal plane mask ($4.3\lambda/D_{apod}$), this configuration provides enhance robustness to tip/tilt, stellar diameter and low-order aberrations\cite{ndiaye2015ApJ_aplc4, ndiaye2016}. The Lyot stop outer diameter is undersized by a factor 0.83 from the apodizer circumscribed aperture, so that the final focal plane IWA and OWA are $3.11\lambda/D_{lyot}$ and $12.45\lambda/D_{lyot}$, as illustrated in Figure \ref{fig:aplc_contrast_profile}. The presented HiCAT APLC has been optimized to increase robustness to the Lyot stop alignment (this solution, 0.1\% alignment robustness).  More recent investigations have shown that the APLC robustness can reach 0.6--1.0\% by combining apodizer robustness optimization for misaligned Lyot stops, and two-DM wavefront control, and the apodizer design will be updated in subsequent fabrication runs\cite{fogarty2018}.

\subsection{Hardware upgrades and metrology}
\begin{figure}[th!]
\includegraphics[width=1.0\textwidth]{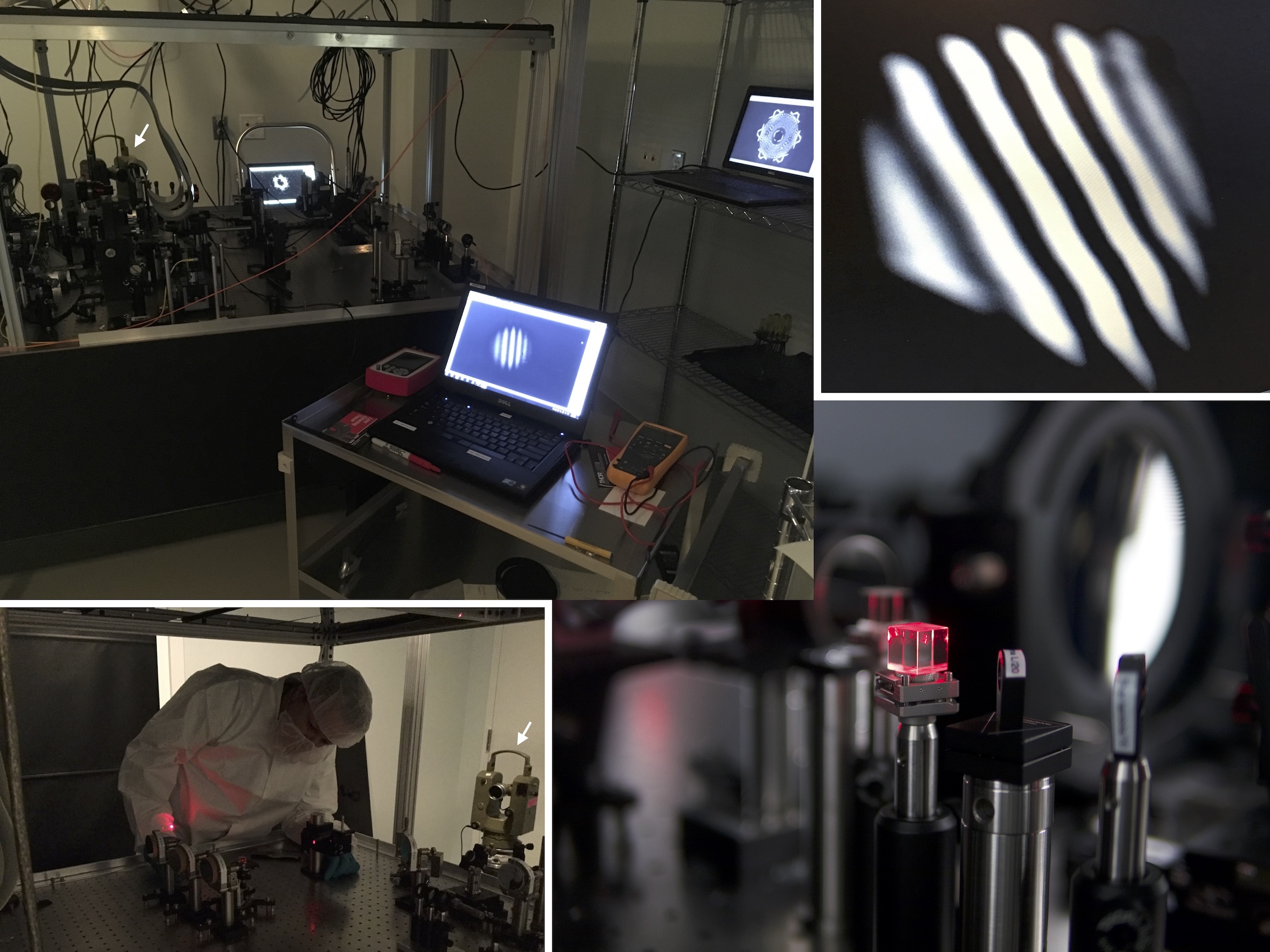}
\caption{Illustration of the HiCAT hardware metrology process for fast replacement of key optics (the apodizer and the three deformable mirrors). Theodolites (indicated by small white arrows) are used to register the tip-tilt of the test optic; Michelson interferometers (see cube, bottom right, and white-light fringes on laptop and insert) are used to register the location along the optical axis; the pupil camera is used for lateral translation alignment; the focal coronagraphic camera allows to assess and check the overall quality. The top figure shows the theodolite, and three laptop screens (imaging camera, pupil camera, and Michelson camera). This process has been used several times successfully either for switching apodizer, for removing and re-installing the DMs for calibration purposes, and to alternate between segmented and non-segmented configurations, as we develop the software infrastructure.  
\label{fig:metrology_process} }
\end{figure}

Over the past year, we have completed a number of hardware upgrades, in particular the installation of three critical components: the 37-segment Iris AO DM, the second Boston kilo-DM, and a series of apodizer. We relied on a suite of external metrology systems, including theodolites, several imaging cameras, and custom white-light Michelson interferometers to enable fast, precise, and repeatable replacement of these critical components. Most components can be swapped safely in just a few hours. This process is illustrated in figure \ref{fig:metrology_process}.

\begin{figure}[th!]
\includegraphics[width=1.0\textwidth]{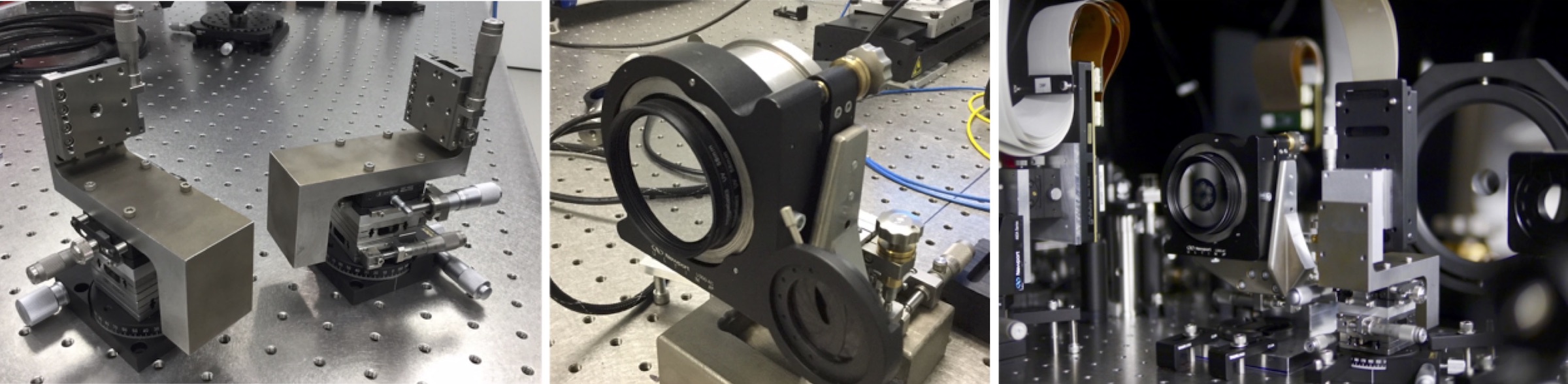}
\caption{Recent HiCAT hardware upgrade developed by a team of Johns Hopkins University undergraduate students. Left: deformable mirror mounts, with 5 degrees of freedom and center of mass balancing. Center: combined apodizer and piezo tip-tilt mount. The apodizers are mounted in a cell based on cheap off-the-shelf photographic camera filters, which allows for safe handling and easy replacement and storage of fragile apodizers. Right: view of apodizer in tip/tilt mount and the three DMs on the HiCAT table.
\label{fig:jhu_hardware} }
\end{figure}

To aid in rapid development, we have set up an adjacent metrology testbed dedicated to the integration and calibration of key hardware components. For example, it was used for developing and validating our external metrology using Michelson interferometers and theodolites, and validating the apodizer swap process using a coordinate measurement arm. We continue to use this metrology testbed for testing new components such as the new mounts developed by a team of Johns Hopkins University Mechanical Engineering undergraduate students (Figure \ref{fig:jhu_hardware}), and for the optimization of the tip-tilt closed loop control parameters. In the near-future we plan to use this testbed for developing an automated target acquisition, and a low-order wavefront sensor\cite{ndiaye2014SPIE_zelda,2016A&A...592A..79N}. 

\subsection{Software development and process}

The HiCAT software was redesigned from the ground up as a modern, configurable, properly-architected software package\cite{moriarty2018}. An experiment queue fully automates execution of coronagraphic experiments and calibration data collections, typically run overnight to preserve physical access to the testbed during the day. A safety system monitors environmental sensors to protect the power- and humidity-sensitive DMs. All data are uniformly processed by a real-time reduction pipeline, also recording the DM commands and configuration parameters. Algorithm development is supported by an integrated testbed software simulator that allows running testbed controls separately from the hardware and generating model images.

This combination of hardware process and advanced software infrastructure has allowed us to perform several major hardware rotations in the past few months, including installations of the second Boston DM and the Iris-AO DM, and tests of three different apodizers (a black-silicon WFIRST apodizer provided by JPL, and the two nanotube apodizers shown in Figure \ref{fig:apodizers}). Each of these five interventions were completed in less than a day, including testbed re-alignment under external metrology, complete re-calibration using our calibration software suite, and the automatic re-creation of a dark zone via overnight unattended operations.

\subsection{Wavefront calibration }

\subsubsection{Parametric Phase Retrieval}
\begin{figure}[t!]
\centering
\includegraphics[width=0.8\textwidth]{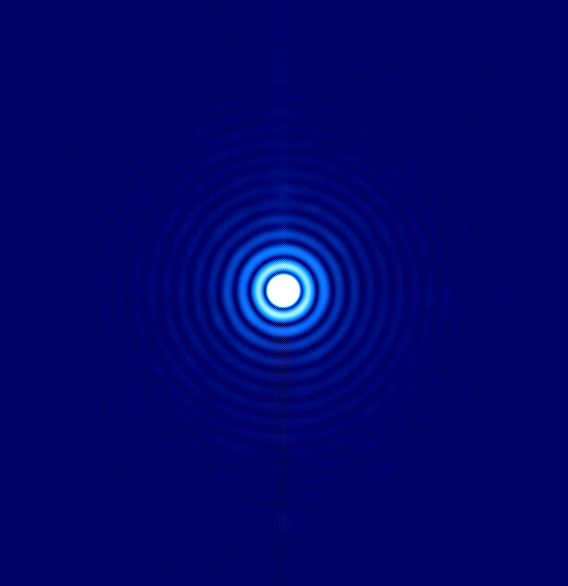}
\caption{PSF at the phase retrieval camera, as a proxy for the PSF at the focal plane mask, except for one reflection on a high-quality mirror to minimize non-common path errors (estimated to be less than 1nm rms). This PSF is obtained for a 18mm pupil diameter, and after 6 off-axis parabolas, 2 toric mirrors, the two Boston DMs and 2 additional flats (total of 13 reflections). This Log scale representation shows a nearly perfect Airy pattern (satellite diffraction spots from DM print through are outside the field of view shown here). Using parametric phase retrieval we have calibrated the residual wavefront error to 3 nm rms or less.  
\label{fig:psf} }
\end{figure}

We use the phase retrieval arm shown in Figure \ref{fig:hicat_cad} to acquire data for a parametric phase retrieval analysis\cite{brady2018}. This phase retrieval approach consists in searching optimal model parameter values that minimize an objective function, here a distance between the model and the data.  This approach allows both for a modal estimation (e.g. using Zernike polynomials) or a pixel-by-pixel direct estimation. We used this analysis for compensating the testbed optical alignment error at the focal plane mask, using DM control.  For that, we placed the testbed in a full circular pupil configuration (i.e. without apodizer and segmented mirror) corresponding to the entire DM aperture (the apodizer clear aperture is slightly undersized compared to the DM for alignment and Fresnel propagation purposes). In this configuration the testbed wavefront error was 16 nm rms with both DMs controlled to their calibrated flat maps. Using two iterations of parametric phase retrieval we have brought the wavefront error to 3 nm rms (Figure \ref{fig:psf})

\subsubsection{COFFEE}
 
\begin{figure}[t!]
\centering
\includegraphics[width=1.0\textwidth]{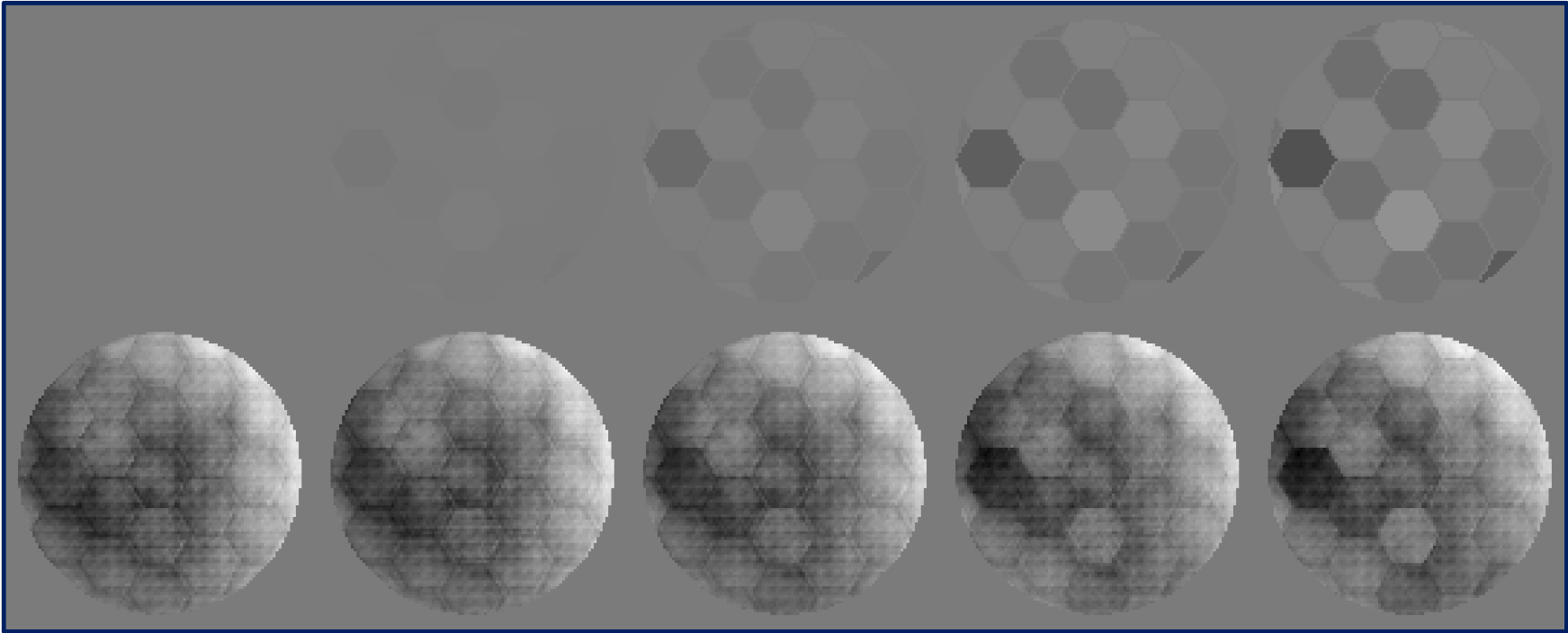}
\caption{COronagraphic Focal-plane wave-Front Estimation for Exoplanet detection (COFFEE) wavefront estimation using sensing through the coronagraph (i.e. after the focal plane mask and Lyot stop), in the presence of the IRIS-AO segmented DM. This is the first demonstration of COFFEE with a segmented aperture. In this reconstruction, we introduced known pistons ramps on the DM (top row), and successfully recovered the wavefront pattern using COFFEE (bottom row). Note that the high spatial frequency print through is modelled by COFFEE, as well as segment discontinuities, without prior information to the algorithm. The successive wavefront errors of the injected phase ramps are, from left to right: 0 (well-phased), 1.9 nm rms, 7.5 nm rms, 13nm rms, 19 nm rms. COFFEE provides a full model of both the upstream and downstream fields (before/after the coronagraph) and is therefore well-suited to be used for high-contrast wavefront sensing for dark zone control on HiCAT in the future.
\label{fig:coffee} }
\end{figure}

The COronagraphic Focal-plane wave-Front Estimation for Exoplanet detection (COFFEE) technique is an extension of the phase diversity technique to coronagraphic images\cite{2012SPIE.8446E..8BP,2013A&A...552A..48P}.
Two coronagraphic images, $I_f$ and $I_d$, are needed. $I_d$ is obtained after applying a known diversity phase $\phi_d$ in the entrance pupil plane, while $I_f$ is on focus. COFFEE enables to estimate two unknown phase aberrations: $\phi_{up}$ corresponds to the upstream phase aberrations, ie. before the coronagraph (typically in the apodizer plane) and $\phi_{do}$ to the downstream aberrations, ie. after the coronagraph (typically in the Lyot stop plane).
A criterion based on the differences between images issued from the coronagraph model and real images has to be minimized. It usually calls a numerical minimization based on the Variable Metric with Limited Memory and Bounds (VMLM-B) method\cite{2014A&A...572A..32P}. The minimization stops when the difference between two successive values of the criterion is below a certain threshold, fixed by the user.
We have successfully tested COFFEE on HiCAT in a configuration with a circular pupil mask and the segmented aperture DM, but without the apodizer in place (i.e. in a ``classical Lyot'' configuration).  By introducing known ramps of pistons on the deformable mirror to study the performance of the algorithm (see Figure \ref{fig:coffee}). In this first test of COFFEE with a segmented aperture, we find that we are able to reconstruct the wavefront to a high-precision from data taken in coronagraphic mode.  We will have to extend next this result to the case of segmented and apodizer apertures. Since COFFEE reconstructs the amplitude and phase at the pupil plane prior to the coronagraph, it is well-suited as a wavefront sensor for dark zone control. It only relies on small defocus introduced on the DM, which we have calibrated in closed-loop with our Fizeau interferometer. This makes COFFEE potentially more advantageous than a probe-based approach such as EFC for which the probes are harder to calibrate. 
 
\section{Conclusions and perspectives}

The main infrastructure of the HiCAT testbed is now completed, with all hardware integrated on the bench. We have designed and manufactured the first carbon nanotubes apodizer and implemented the first APLC for a truly segmented aperture. Our first dark zones using single-DM speckle nulling ($1.7\times10^{-6}$ monochromatic and $6.3\times10^{-6}$ in $6\%$ broadband) demonstrate this technology at TRL3. Our software infrastructure and hardware metrology and process allows high flexibility (hardware configuration changes, autonomous unattended software operations). We are in the process of integrating a two-DM wavefront control\cite{pueyo2009ApOpt,2018AJ....155....7M,2018AJ....155....8M}. In the near future, we plane to continue improving the overall infrastructure and performance, including tip-tilt control and low-order wavefront control. 

\acknowledgments 
This work was supported in part by the National Aeronautics and Space Administration under Grants NNX12AG05G and NNX14AD33G issued through the Astrophysics Research and Analysis (APRA) program (PI: R. Soummer) and by Jet Propulsion Laboratory subcontract No.1539872 (Segmented-Aperture Coronagraph Design and Analysis; PI: R. Soummer), and the STScI Director's Discretionary Research Funds. It is also partly funded by the French national aerospace research center ONERA (Office National d’Etudes et Recherches Aérospatiales) and by the Laboratoire d'Astrophysique de Marseille (LAM). We are also thankful to Lee Feinberg and GSFC for lending us theodolites and a coordinate measurement Faro arm, and thankful to Stuart Shaklan from JPL for loaning us a WFIRST-CGI apodizer, which helped us accelerate the development of our software and hardware infrastructure. 
\bibliography{bibliotheque}
\bibliographystyle{spiebib}

\end{document}